\documentclass[%
aip,
jcp,
 amsmath,amssymb,
reprint,
]{revtex4-1}

\usepackage{graphicx}
\usepackage{dcolumn}
\usepackage{bm}

\usepackage{amsmath}	
\begin{document}

\preprint{AIP/123-QED}

\title[Note: Control parameter dependence of fluctuations near jamming]
{Note: Control parameter dependence of fluctuations near jamming}
\author{Harukuni Ikeda}
 \email{harukuni.ikeda@gakushuin.ac.jp}
\affiliation{
Department of Physics, Gakushuin University, 1-5-1 Mejiro, Toshima-ku, Tokyo 171-8588, Japan}


\date{\today}

\maketitle

\newcommand{\diff}[2]{\frac{d#1}{d#2}}
\newcommand{\pdiff}[2]{\frac{\partial #1}{\partial #2}}
\newcommand{\fdiff}[2]{\frac{\delta #1}{\delta #2}}
\newcommand{\bx}{\bm{x}}
\newcommand{\br}{\bm{r}}
\newcommand{\ba}{\bm{a}}
\newcommand{\by}{\bm{y}}
\newcommand{\bY}{\bm{Y}}
\newcommand{\bF}{\bm{F}}
\newcommand{\bn}{\bm{n}}
\newcommand{\be}{\bm{e}}
\newcommand{\new}{\nonumber\\}
\newcommand{\abs}[1]{\left|#1\right|}
\newcommand{\tr}{{\rm Tr}}
\newcommand{\HH}{{\mathcal H}}
\newcommand{\OO}{{\mathcal O}}
\newcommand{\var}{{\rm Var}}
\newcommand{\ave}{{\rm Ave}}


In this work, we report the control parameter dependence of the
fluctuations near the jamming transition point. We show that the
fluctuations do not diverge in pressure control, while it diverges in
packing fraction control.

We consider purely repulsive harmonic discs in a two-dimensional
$L\times L$ box with periodic boundary conditions at zero
temperature~\cite{ohern2003}:
\begin{align}
& V_N = \sum_{i<j}^{1,N}\frac{h_{ij}^2}{2}\theta(-h_{ij}),\ h_{ij} = \abs{\br_i-\br_j}-R_i-R_j,\
\end{align}
where $N$ denotes the number of particles, $\br_i = (x_i, y_i)$ denotes
the position, and $R_i$ denotes the radius. To avoid crystallization, we
consider a $50:50$ binary mixture of large $R_L=0.7$ and small $R_s=0.5$
particles. The value of $V_N$ separates the jammed and unjammed phases:
when the packing fraction $\varphi=N\pi(R_s^2+R_L^2)/(2L^2)$ is smaller
than the jamming transition point $\varphi_J$, one observes $V_N=0$
after the energy minimization, while, when $\varphi>\varphi_J$, $V_N$
has a finite value. For the energy minimization, we use the fast
inertial relaxation engine (FIRE)~\cite{fire2006}. We terminate the
energy minimization when $\sum_{i=1}^N
(\partial_{\br_i}V_N)^2/N<10^{-25}$. In our numerical simulation, we
define $\varphi_J$ at which the energy barely has a finite value
$V_N/N\in (10^{-16},2\times 10^{-16})$ after the energy minimization. We
generate the configurations above $\varphi_J$ in two ways, as described
below.
\paragraph{Packing fraction control}
We use $\varepsilon=\varphi-\varphi_J$ as a control parameter. Following
O'\ Hern \textit{et al.}, we first generate the configuration at
$\varphi_J$ by combining compression and decompression: we compress the
system when $V_N<10^{-16}$ and decompress when $V_N>10^{-16}$, see
Ref~\cite{ohern2003} for details. After every compression/decompression,
we minimize the energy by using the FIRE algorithm~\cite{fire2006}. We
terminate the process when $V_N/N\in (10^{-16},2\times 10^{-16})$. After
obtaining a configuration at $\varphi_J$, we re-compress as the amount
of $\varepsilon=\varphi-\varphi_J$ to obtain a configuration above
jamming. As reported in Ref.~\cite{vanderwerf2020}, some samples unjam
after the compression (compression unjamming). We throw out such
samples.

\paragraph{Pressure control}
The pressure $p$ is used as the control parameter. For this purpose, we
repeat the compression and decompression until the system's pressure
reaches the target pressure. In this case, the jamming transition point
corresponds to $p=0$. 

For each $\varepsilon$ and $p$, we prepare $M=1000$ independent samples
and calculate the mean and variance of physical quantities.
We only use the data for $p\gg 10^{-6}$ so that the
force balance tolerance and energy tolerance do not affect the results.

\begin{figure}[t]
 \begin{center}
\includegraphics[width=9cm]{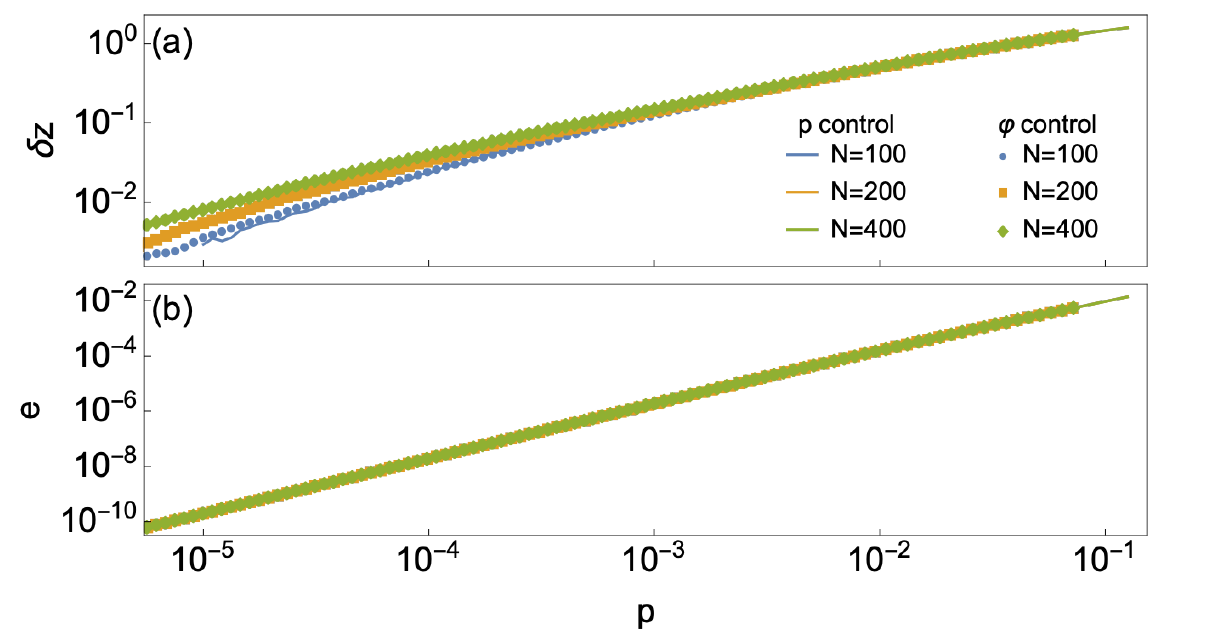} 
\vspace{-0.8cm}
 \caption{ Fluctuations of physical quantities. Solid lines and markers
denote data for $p$ and $\varphi$ control, respectively. For comparison,
results for $\varphi$ control are plotted as a function of the average
$p$ at each $\varphi$. (a) Contract number. (b) Energy per particle $e=V_N/N$.}
\label{204406_17Jul22}
 \end{center}
 \vspace{-0.8cm}
\end{figure}

We first discuss the behaviors of the average quantities. A commonly
observed quantity to characterize the jamming transition is the number
of contacts per particle $z$. At $\varphi_J$, $z$ converges to $z_J=
2d-2d/N+2/N$, if one removes rattlers that have less than three
contacts~\cite{PhysRevLett.109.095704}. Hereafter we remove the rattlers
when calculating $z$. Another commonly used quantity is the energy per
particle $e=V_N/N$. In Fig.~\ref{204406_17Jul22}, we plot the average
values of the excess contact number $\delta z = z-z_J$ and $e$. It can be
seen that the average values do not depend on the control parameters.

\begin{figure}[t]
\begin{center}
 \includegraphics[width=9cm]{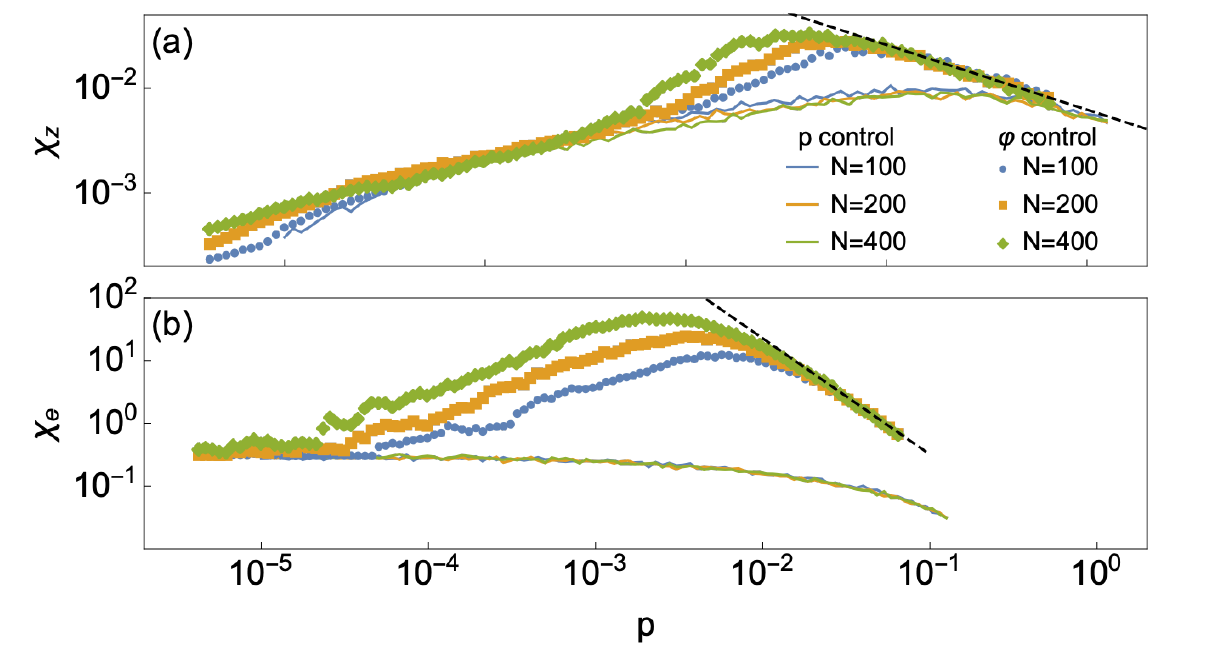} 
\vspace{-0.8cm}
\caption{Fluctuations of physical quantities. Solid lines and markers
 denote data for $p$ and $\varphi$ control, respectively. Black dashed
 lines denote power-law fit. For comparison, results for $\varphi$
 control are plotted as a function of the average $p$ at each
 $\varphi$. (a) Fluctuation of the contact number. (b) Fluctuation of
 energy.}  \label{225318_17Jul22}
\end{center}
 \vspace{-0.8cm}
\end{figure}
Next, we discuss the fluctuations.
To see how large the fluctuation
is compared to the mean value, we observe the following quantities:
\begin{align}
 \chi_{z} \equiv N\frac{{\rm Var}(z)}{{\rm Ave}(z)^2},\
 \chi_{e} \equiv N\frac{{\rm Var}(e)}{{\rm Ave}(e)^2},
\end{align}
where ${\rm Var}$ and ${\rm Ave}$ respectively denote the variance and
average for the $M$ samples. The factor $N$ guarantees that $\chi_{z,e}$
converges to a finite value in the thermodynamic limit
$N\to\infty$~\cite{PhysRevLett.123.068003}. In
Fig.~\ref{225318_17Jul22}, we plot the numerical results of $\chi_z$ and
$\chi_e$. In $p$ control, the $N$ dependence of $\chi_z$ only appears
very near the jamming transition point, $p\lesssim 10^{-4}$. The finite
size effects in this region are examined in detail in
Ref.~\cite{PhysRevLett.123.068003}. We do not observe any significant
$N$ dependence for $\chi_e$. On the contrary, in $\varphi$ control, both
$\chi_z$ and $\chi_e$ significantly increase with $N$ in the
intermediate region ($10^{-3}\lesssim p\lesssim 10^{-2}$ for $\chi_z$,
and $10^{-5}\lesssim p\lesssim 10^{-2}$ for $\chi_e$). In the
intermediate $\varphi$ region, $\chi_{z,e}$ is well fitted with the
power-law function:
\begin{align}
\chi_{z,e} = A_{z,e}p^{-\beta_{z,e}},
\end{align}
where $\beta_z=0.62$ and $\beta_e=1.85$, see black dashed lines in
Fig.~\ref{225318_17Jul22}. The power-law region increases with $N$, and
in the thermodynamic limit, the fluctuations are expected to diverge at
the transition point.
\begin{figure}[t]
\begin{center}
 \includegraphics[width=9cm]{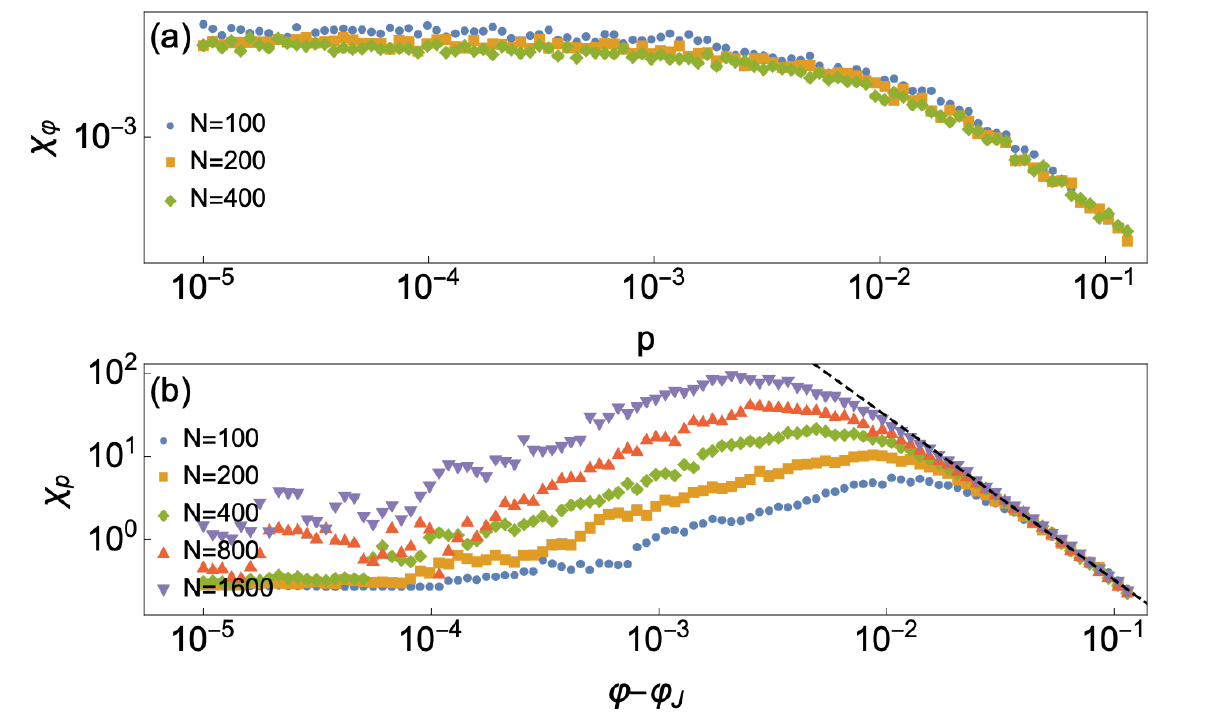} 
\vspace{-0.8cm}
\caption{Fluctuations of (a) $\varphi$ in $p$ control and (b) $p$ in
 $\varphi$ control. Markers denote numerical results, while the dashed
 line shows the power-law fit.}  \label{171853_21Jul22}
\end{center}
 \vspace{-0.8cm}
\end{figure}
In Fig.~\ref{171853_21Jul22}, we plot the fluctuation of $\varphi$,
${\chi_\varphi \equiv N\var(\varphi)/\ave(\varphi)^2}$, in $p$
control and the fluctuation of $p$, ${\chi_p\equiv
N\var(p)/\ave(p)^2}$, in $\varphi$ control. We found that
$\chi_\varphi$ remains finite, while $\chi_p$ exhibits a power-law
divergence ${\chi_p\sim (\varphi-\varphi_J)^{-\beta_p}}$ with $\beta_p =
1.97$, see the dashed line in Fig.~\ref{171853_21Jul22}.

\begin{figure}[t]
\begin{center}
 \includegraphics[width=9cm]{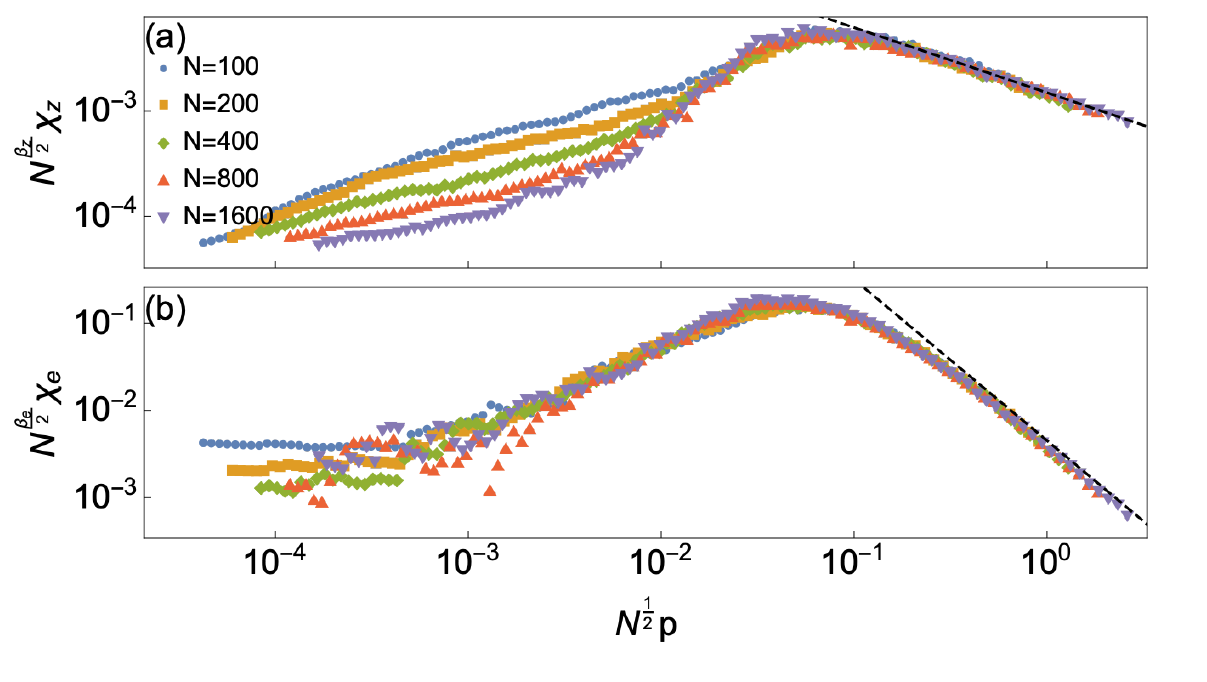}
\vspace{-0.8cm}
\caption{Scaling plots for (a) $\chi_z$ and $(b)$ $\chi_e$ in $\varphi$
control. Markers denote numerical results, while the dashed lines denote
power-law fit $A_{z,e}p^{\beta_{z,e}}$.} \label{121601_24Nov22}
\end{center}
\vspace{-0.8cm}
\end{figure}

Finally, we propose a phenomenological model to explain the divergence
of the physical quantities in $\varphi$ control.
Fig.~\ref{171853_21Jul22}~(a) and a previous research~\cite{ohern2003}
show that the variance of $\varphi$ remains finite at $\varphi_J$. Also,
$p\propto \varphi-\varphi_J$ near
$\varphi_J$~\cite{ohern2003}. Therefore, $p$ and $\varphi$ have the
following linear relation near $\varphi_J$: $\varphi=\varphi_J+ Ap+
\xi$, where $\xi$ is a random variable of zero mean and variance
$\overline{\xi^2}=\Delta/N$~\footnote{F.~Zamponi (private
communication)}, and $A$ and $\Delta$ are constants. Then, $p$ can be
expressed as a function of $\delta\varphi$:
\begin{align}
p=A^{-1}(\delta\varphi-\xi),\label{115546_25Nov22}
\end{align}
with $\delta\varphi=\varphi-\varphi_J$. It is straightforward to
show $\chi_p\sim \delta\varphi^{-\beta_p^{th}}\sim p^{-\beta_p^{th}}$
with $\beta_p^{th}=2$, which is close to the numerical result
$\beta_p=1.97$. Since the energy is a quadratic function of $p$
~\cite{ohern2003}, $e\sim p^2 = A^{-2}(\delta\varphi^2 -2\delta\varphi
\xi+\cdots)$, leading to $\chi_e\sim p^{-\beta_e^{th}}$ with
$\beta_e^{th}=2$. Again, this is close to the numerical result
$\beta_e=1.85$. The contact number exhibits the square-root singularity
$z-z_J\sim p^{1/2}
=A^{-1/2}\left(\delta\varphi^{1/2} - \xi
\delta\varphi^{-1/2}/2+\cdots\right)$, leading to $\chi_z\sim
p^{-\beta_z^{th}}$ with $\beta_z^{th}=1/2$, which is slightly underestimated
but close to the numerical result $\beta_z=0.62$. The above mean-field
like argument may no-longer hold when the fluctuation of the pressure
$A^{-1}\xi$ becomes larger than the mean-value $A^{-1}\delta\varphi$ in
Eq.~(\ref{115546_25Nov22}), which defines the characteristic pressure
$p\sim\delta\varphi\sim\xi\sim O(N^{-1/2})$. This consideration suggests
the following scaling form:
\begin{align}
\chi_{z,e} = N^{\frac{\beta_{z,e}}{2}}f_{z,e}(N^{\frac{1}{2}}p),
\end{align}
where $f_{z,e}(x)$ denotes the scaling function such that
${f_{z,e}(x)\sim x^{-\beta_{z,e}}}$ for $x\ll 1$. In
Fig.~\ref{121601_24Nov22}, we confirmed the above scaling ansatz using
the data shown in Fig.~\ref{225318_17Jul22} and data for larger $N$ in
$\varphi$ control.

\vspace{-0.5cm}
\acknowledgments
\vspace{-0.5cm}
We thank A.~Ikeda, P.~Urbani, and F.~Zamponi for useful comments. This work was
supported by KAKENHI 21K20355.
\vspace{-0.7cm}
\subsection*{DATA AVAILABILITY} 
\vspace{-0.5cm}
The data that support the findings of this study are available from the
corresponding author upon reasonable request.

\bibliography{reference}

\end{document}